\documentclass[english,aps,two column]{revtex4}
\usepackage[T1]{fontenc}
\usepackage[latin1]{inputenc}
\usepackage{graphicx}
\usepackage{amssymb}

\makeatletter

 \newenvironment{lyxlist}[1]
   {\begin{list}{}
     {\settowidth{\labelwidth}{#1}
      \setlength{\leftmargin}{\labelwidth}
      \addtolength{\leftmargin}{\labelsep}
      }}
   {\end{list}}

\usepackage{babel}
\makeatother
\begin{document}

\title{Effect of the Triplet State on the Random Telegraph Signal in Si
n-MOSFETs}

\author{Enrico Prati}

\affiliation{Laboratorio Nazionale Materiali e Dispositivi per la Microelettronica,
Consiglio Nazionale delle Ricerche - Istituto Nazionale per la Fisica
della Materia, Via Olivetti 2, I-20041 Agrate Brianza, Italy}

\author{Marco Fanciulli}

\email{marco.fanciulli@mdm.infm.it}

\affiliation{Laboratorio Nazionale Materiali e Dispositivi per la Microelettronica,
Consiglio Nazionale delle Ricerche - Istituto Nazionale per la Fisica
della Materia, Via Olivetti 2, I-20041 Agrate Brianza, Italy}

\author{Giorgio Ferrari}

\affiliation{Dipartimento di elettronica e informazione, Politecnico di Milano,
P.za Leonardo da Vinci 32, I-20133 Milano, Italy}

\author{Marco Sampietro}

\affiliation{Dipartimento di elettronica e informazione, Politecnico di Milano,
P.za Leonardo da Vinci 32, I-20133 Milano, Italy}

\begin{abstract}
We report on the static magnetic field dependence of the random telegraph
signal (RTS) in a submicrometer silicon n-metal-oxide-semiconductor
field-effect transistor. Using intense magnetic fields and $^{3}$He
temperatures, we find that the characteristic time ratio changes by
3 orders of magnitude when the field increases from 0 to 12 T. Similar
behaviour is found when the static field is either in-plane or perpendicular
to the two dimensional electron gas. The experimental data can be
explained by considering a model which includes the triplet state
of the trapping center and the polarization of the channel electron
gas. 
\end{abstract}
\maketitle
Random telegraph signal (RTS), consisting of the random switching
of the current in a metal-oxide-semiconductor field effect transistor
between two levels, has been studied for decades \cite{Ralls 84,Kandiah 89,Longoni 95}.
The quantum theory of the RTS in silicon metal-oxide-semiconductor
field-effect transistors (MOSFETs) is based on inelastic capture and
emission of charge by a trapping center near the $Si/SiO_{2}$ interface.
Each tunnel transition of the electron between the trap and the channel
is assisted by multiphonon emission and absorption, and reflects the
effect of electrostatic Coulomb barriers of the charges present in
the device \cite{Palma 97}. 

Recently RTS has been proposed as a viable way towards single spin
detection, a key issue towards solid state quantum information processing
(QIP) \cite{Kane 98,Loss 98}. In particular, for those QIP schemes
in which the electron spins are the qu-bits, read-out of the spin
state can be achieved by monitoring the variation of the RTS in electron
spin resonance conditions. In this scheme the MOSFET is operated in
a static magnetic field and under microwave irradiation \cite{Vrijen,Mozyrsky 2001,Xiao 04}.
Hence, the response of a trap at the $Si/SiO_{2}$ interface of a
MOSFET placed in a static magnetic field and the occupation of the
trap spin states are of considerable interest. 

The effect of the static magnetic field on the RTS has been recently
described \cite{Xiao 03}, and a model has been proposed to take into
account the characteristic time ratio $r=\frac{\tau_{high}}{\tau_{low}}$
change as a function of the magnetic field, where $\tau_{high}$ and
$\tau_{low}$ are the average characteristic times of the two current
levels. However, at very low temperatures and high magnetic fields,
significant discrepancies between the experimental data obtained and
the model have been observed. 

We report on the investigation of the RTS parameters as a function
of static magnetic fields up to 12 T, both in-plane and perpendicular
to the two dimensional electron gas (2DEG) localised in the inversion
layer of a Si n-MOSFET. Our experiments are performed at liquid $^{3}$He
temperature where the effects due to the static magnetic field are
expected and proved to be strong \cite{Xiao 03}. We also suggest
a model which, taking into account the triplet state of the trap,
can explain the high field effects.

The investigated devices are commercial n-channel MOSFETs realised
on a p-well, with a length of 0.18 $\mu$m and a width of 0.28 $\mu$m,
3.5 nm thick gate oxide, and a threshold voltage at low temperature
$V{}_{\textrm{th}}$= 536 mV. The current $I{}_{D}$ flowing through
drain and source is measured by a transimpedance amplifier whose output
is sampled and digitized for off-line processing. The bandwidth of
the amplifier extends from 0.1 Hz to about 160 kHz. The transistor
and the electronic are powered by independent batteries to avoid power-line
pick-ups and interferences. Extreme caution has been devoted to the
suppression of all the spurious signals due to the instruments. The
setup allowed us to characterize traps with characteristic times down
to few $\mu$s. Each acquisition was taken in a time interval between
60 and 150 s. The samples were mounted on a gold plated dual-in-line
made with high thermal conductivity ceramic placed in thermal contact
with a copper coldfinger. All the contacts to the source, drain, and
gate were directly accessible through bonding pads and connected to
the dual in line pads. The coldfinger was cooled at $^{3}$He cryogenic
temperature in vacuum at the nominal value of $T{}_{cf}$= 245 mK.
The cryostat was operated into a 12 T Oxford Instruments magnet. The
temperature $\Theta$ of the electron gas was slightly higher than
the base temperature $T{}_{cf}$ because of the power dissipation
of the current $I{}_{D}$, observed as a broadening of the single
state resonant tunneling conductance peaks versus the gate voltage
$V{}_{G}$ \cite{Stone 85}. 

Suitable traps for the RTS detection were identified. Each sample
revealed traps at several gate voltages above the threshold and activated
in a voltage range $\bigtriangleup V{}_{G}\simeq$ 0.5 - 2 mV. Characteristic
times of the traps ranged from tens of $\mu$s to few ms.

Figure 1 shows the characteristic time ratio \textit{}$r$ for a trap
observed with $V{}_{G}$= 560.7 mV and $V{}_{D}$= 15.7 mV as a function
of an external static magnetic field $B\Vert$ in the plane of the
channel and oriented in the direction perpendicular to the electron
current flow (inset of Fig. 1). The drain current was $I{}_{D}$=
37.3 nA. The current state \textit{low} of the RTS corresponds to
the physical state of an electron captured by the trap, because the
ratio \textit{r} decreases by increasing $V{}_{G}$ (therefore lowering
the trap energy $E{}_{T}$). 

In order to determine whether the trap was singly occupied, and therefore
paramagnetic, and capable of capturing a second electron (type 1$\rightarrow$
2) or an empty trap capable of capturing an electron (0 $\rightarrow$1),
we have analysed the dependence of \textit{r} on the magnetic field
B \cite{Xiao 04}. For a singly occupied trap, the highest energy
level of the Zeeman doublet increases as the magnetic field increases.
We observed that the trap under investigation increases its permanence
in the current state \textit{high} by increasing the magnetic field,
so that the highest energy level becomes on the average less filled,
implying that the trap is a 1$\rightarrow$ 2 type. Such experimental
fact is consistent with the most probable nature of the trap being
a $P{}_{b0}$ or $P{}_{b1}$ at the $Si/SiO_{2}$ interface or an
$E'$ center close to it. The ratio \textit{r} increases by three
orders of magnitude when the magnetic field increases from 0 to 12
T. Such behaviour is qualitatively similar to that observed by Xiao
et al.\cite{Xiao 03}. However, quantitatively, the effect we detect
is two orders of magnitude larger. 

The same sample was examinated also with the external static magnetic
field $B\bot$ perpendicular to the channel. Figure 2 shows the ratio
\textit{r} for another trap observed with $V{}_{G}$= 673.8 mV and
$V{}_{D}$= 1.6 mV. At this gate voltage, the channel electron density
is $n{}_{s}\simeq8\times10^{11}$ cm$^{-2}$. The current was $I{}_{D}$=
35 nA. This trap has also been investigated with $B\Vert$ as discussed
later. By applying the same procedure than in the previous case, we
found that the active trap involves a 1$\rightarrow$ 2 process. Also
in this case \textit{r} increases by more than a factor 10 when the
magnetic field $B\bot$ was increased from 0 to 11 T (Figure 2). A
peculiar signature of the measurement is that the RTS disappears when
$B\bot$ is in the range 7.75 T and 9.4 T. $\Delta I$ is defined
as the difference between the \textit{high} and \textit{low} state
current and it vanishes in such field range (Fig. 3). The magneto-induced
localization of electrons occurring into the 2DEG at such high magnetic
fields can be responsible of such RTS suppression. At low temperature,
when the 2DEG is orthogonal to the magnetic field B, the energy levels
of the electron gas are discrete (Landau levels) and we point out
that the filling factor $\nu=4$ level is at 8.3 T. The Ramo's theorem
states that $\Delta I=qvE_{x}$ where $q$ is the electron charge,
$v$ the charge velocity component in the direction of the electric
field, and $E_{x}$ the applied electric field. We speculate that
such vanishing of $\Delta I$ reflects the average value $v=0$ due
to localization.

To explain the dependence of $r$ on the magnetic field, we propose
a model based on elementary statistical physics that includes the
triplet state of the trap and the Zeeman splitting of the channel
electrons in addition to the Zeeman splitting of the trap. The model
relies on the following assumptions. The effect of the magnetic field
is to split the spin degenerate levels of conduction electrons (having
g-factor $g{}_{e}$\cite{Wallace 91}) and of the trap (having g-factor
$g{}_{T}$) by $E_{Z}^{c}=g{}_{e}\mu_{B}B$ and $E_{Z}^{T}=g{}_{T}\mu_{B}B$
respectively, where $\mu_{B}=\frac{e\hbar}{2m_{e}c}$. Assuming that
at high field a two-particle triplet state can be formed, we call
$\Delta U$ the difference between the energy in the singlet $S=0$
and in the triplet states $S=1$ of the trap at $B=0$. $\Delta U$
includes both the exchange term and the difference of the Coulomb
repulsion between the electrons in the single and triplet states.
As suggested by Xiao et al.\cite{Xiao 03} the Coulomb respulsion
can be considered as B field independent. We assume that the inelastic
tunnelling of the channel electrons into the trap occurs in two steps:
first, the preparation of the electron to the capture process because
of both the interaction with the potential of the trapping center
and the multiphonon emission due to the relaxation of the lattice
near the defect \cite{Henry 77}; second the capture itself by tunneling
(and viceversa for the emission)\cite{Palma 97}. The temperature
$\bar{\Theta}$ associated to such a process is locally defined and
refers to the inelastically scattered electrons prepared to be captured
and the paramagnetic electron of the trap. The expression of the ratio
\textit{r} is:\[
r=\frac{2e^{\beta(E_{T}+U_{S}-E_{F})}\left(\cosh\beta\left\langle E_{Z}\right\rangle +\cosh\beta\frac{\delta}{2}\right)}{1+e^{-\beta\Delta U}\left(1+\cosh\beta E_{Z}^{T}\right)}(1)\]
where $E_{T}$ is the energy level of the empty trap, $U_{S}$ is
the Coulomb energy of the doubly occupied trap, $\beta=k\bar{\Theta}$
, $k$ is the Boltzmann constant,$\left\langle E_{Z}\right\rangle =\frac{E_{Z}^{T}+E_{Z}^{C}}{2}$
is the average value of the two Zeeman splittings of the channel and
of the trap respectively, and $\delta=\left|E_{Z}^{T}-E_{Z}^{C}\right|$.
Since the g-factors of the free electron and of the defect are very
close (assuming that the trap is a $P{}_{b0}$,$P{}_{b1}$ or an $E'$
center), then $\cosh\beta\frac{\delta}{2}\approx1$. The denominator
differs from unity because of a term which depends on the inclusion
of the triplet state. At high B field such term dominates and suppresses
the exponential behaviour of $r$ as a function of B. Eq. (1) has
been used to fit the experimental data when the B field was either
parallel (Figure 1) or perpendicular (Figure 2) to the channel. The
fit was performed by considering the logarithm of the Eq. 1 to fit
the logarithm of the experimental data so all the points were weigthed
in the same way. The three parameters used for the model which includes
the triplet state were the local temperature $\bar{\Theta}$ of the
electrons interacting with the trap site, a multiplicative constant,
and $\Delta U$, while only the first two parameters were used if
only the $S=0$ state was considered. For both the experiments ($B\Vert$
and $B\bot$ to the 2DEG) a significantly better fit was obtained
by including the triplet state. The improvement of the fit including
the polarization of the 2DEG was less important, but not negligible.
In the $B\Vert$ and $B\bot$ cases, the fit including the triplet
state gave a local temperature $\bar{\Theta}$ of 1.5 K and 2.3 K
respectively, with an error of about 5\%. The second trap has been
measured also applying $B\Vert$ in the same operating conditions.
$B\Vert$ was swept to an induction of 12 T in both the positive and
negative directions. The fit gave the same temperature obtained for
the $B\bot$ case. The triplet state model gives a $\Delta U\approx5k_{B}\bar{\Theta}$
where $\bar{\Theta}$ is the temperature calculated by the same fit
for both traps. At first sight it could be surprising that a higher
temperature is found in the second trap (B field perpendicular to
the 2DEG) where a smaller drain voltage was applied and a smaller
current was flowing through the channel. However, this fact is compatible
with the scheme proposed, because the local electron temperature in
the neighborhood of the defect depends not only on the power dissipated
by the current, but also on the trap distance from the $Si/SiO_{2}$
interface, its proximity to the source along the channel, and its
energy level, via the number of phonons involved in the inelastic
tunneling process.

To conclude, we have demonstrated experimentally that the ratio between
the characteristic times of the RTS in a silicon n-MOSFET may change
by three orders of magnitude in presence of an intense magnetic field
up to 12 T parallel to the 2DEG. The same behaviour, although less
pronounced, was observed also in another trap by applying a perpendicular
field. To explain quantitatively the change of the ratio $r$ as function
of B, and the suppression of the exponential behavior when high magnetic
fields are reached, we propose a model that includes the involvement
of the triplet state of the trap and the polarization of the 2DEG.

\begin{acknowledgments}
The authors would like to thank Ivar Martin (Los Alamos), Dima Mozyrsky
(Clarkson University, NY), and Hong Wen Jiang (UCLA, CA), for the
useful discussion, Vittorio Pellegrini, Vincenzo Piazza, Pasqualantonio
Pingue for the cryomagnet time and the support at the Laboratorio
Polvani of NEST CNR-INFM and SNS, Pisa, Italy.
\end{acknowledgments}

\section{Figure Captions}

\begin{lyxlist}{00.00.0000}
\item [Fig.1]The ratio \textit{r} as a function of $B\Vert$. Because of
the power dissipated by the current, the temperature of the electrons
involved in the RTS of the trap here considered has been evaluated
being 1.5 K, instead of the nominal cold-finger temperature of 245
mK, by fitting the experimental results (black points) with the Eq.
1 (line). Inset: the magnetic field is in the plane of the 2DEG and
perpendicular to the current channel. 
\item [Fig.2]The ratio \textit{r} as a function of $B\bot$. The experimental
measurement (black points) has been fitted by using Eq. 1, obtaining
a local temperature 2.3 K, instead of the nominal cold-finger temperature
of 245 mK. The RTS disappears between 7.75 T and 9.4 T. The inset
shows the direction of the magnetic field, orthogonal to the sample. 
\item [Fig.3]The $\Delta I$ changes as a function of the magnetic field
$B\bot$. $\Delta I$ vanishes between 7.75 T and 9.4 T. This fact
should be related to magneto-induced localization of conduction electrons.
At the electron density of the 2DEG used during the measurement, the
Landau level $\nu=4$ is at 8.3 T.
\end{lyxlist}
Figure 1\includegraphics[%
  scale=0.9]{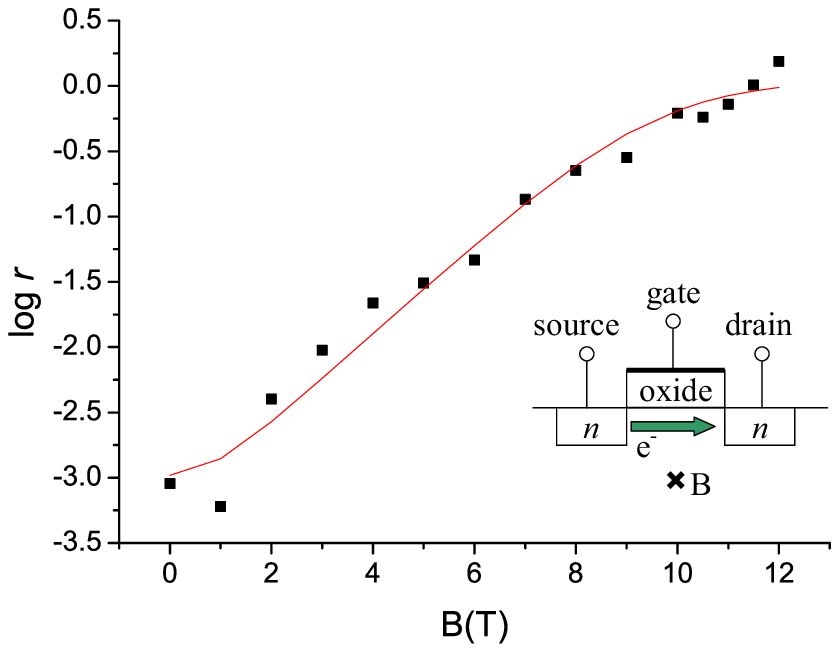}

Figure 2\includegraphics[%
  scale=0.9]{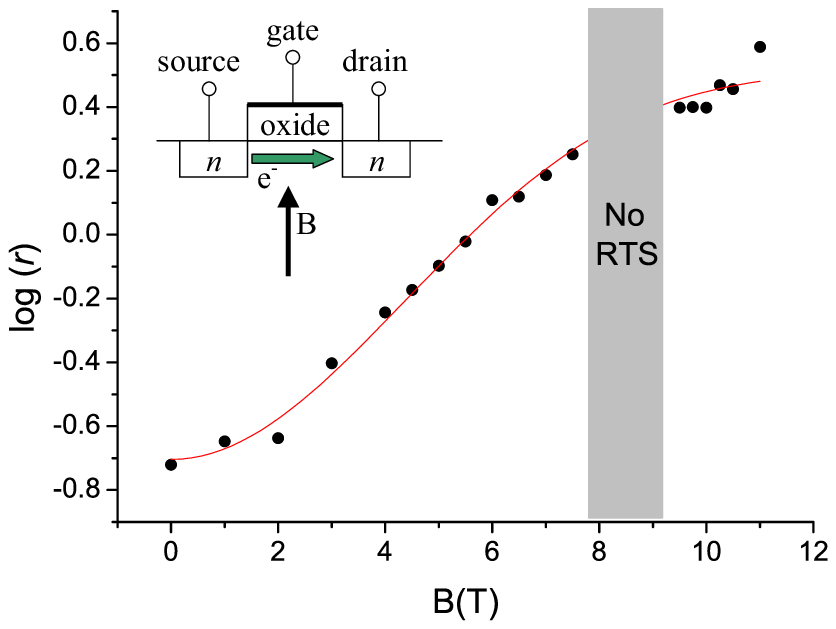}

Figure 3\includegraphics[%
  scale=0.8]{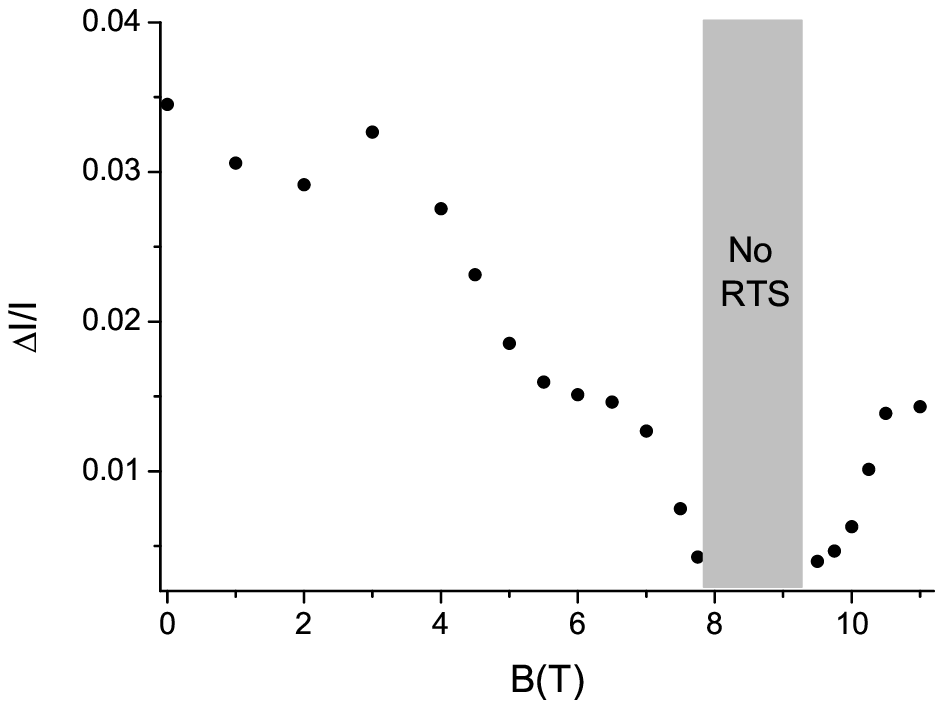}
\end{document}